\documentclass{emulateapj}
\usepackage{graphicx,natbib,aas_macros}

\submitted{Submitted to IAU Symposium 383 --- Astrochemistry VIII: From the First Galaxies to the Formation of Habitable Worlds}

\begin{document}

\title{ALMA Spectroscopy of Europa: A Search for Active Plumes}

\author{Cordiner, M. A.\altaffilmark{1,2}, Thelen, A. E.\altaffilmark{3}, Lai, I.-L\altaffilmark{4}, Tseng, W.-L.\altaffilmark{5}, Nixon, C. A.\altaffilmark{1}, Kuan, Y.-J.\altaffilmark{5,6}, Villanueva, G. L.\altaffilmark{1}, Paganini, L.\altaffilmark{7}, Charnley,~S.~B.\altaffilmark{1}, Retherford, K. D.\altaffilmark{8,9}}

\altaffiltext{1}{ Solar System Exploration Division, NASA Goddard Space Flight Center, 8800 Greenbelt Road, Greenbelt, MD 20771, USA. }
\altaffiltext{2}{ Department of Physics, Catholic University of America, Washington, DC 20064, USA.}
\altaffiltext{3}{ Division of Geological and Planetary Sciences, California Institute of Technology, Pasadena, CA 91125, USA.}
\altaffiltext{4}{ Graduate Institute of Astronomy, National Central University, Taoyuan 320, Taiwan.}
\altaffiltext{5}{ Department of Earth Sciences, National Taiwan Normal University, Taipei 116, Taiwan.}
\altaffiltext{6}{  Institute of Astronomy and Astrophysics, Academia Sinica, Taipei 106, Taiwan.}
\altaffiltext{7}{ NASA Headquarters, Washington, DC 20546, USA.}
\altaffiltext{8}{ Southwest Research Institute, San Antonio, TX 78238, USA.}
\altaffiltext{9}{ University of Texas at San Antonio, San Antonio, TX 78249, USA.}

\begin{abstract}
The subsurface ocean of Europa is a high priority target in the search for extraterrestrial life, but direct investigations are hindered by the presence of a thick, exterior ice shell. Here we present spectral line and continuum maps of Europa obtained over four epochs in May-June 2021 using the Atacama Large Millimeter/submillimeter Array (ALMA), to search for molecular emission from atmospheric plumes, with the aim of investigating subsurface processes. Using a 3D physical model, we obtained upper limits for the plume abundances of HCN, H$_2$CO, SO$_2$ and CH$_3$OH.  If active plume(s) were present, they contained very low abundances of these molecules. Assuming a total gas production rate of $10^{29}$~s$^{-1}$, our H$_2$CO abundance upper limit of $<0.016$\% is more than an order of magnitude less than measured in the Enceladus plume by the Cassini spacecraft, implying a possible chemical difference between the plume source materials for these two icy moons.

\end{abstract}

\section{Introduction}

The population of icy bodies within our solar system includes dwarf planets, comets, Kuiper belt objects and some of the moons of the giant planets. Despite their differing bulk properties and orbits, these bodies share a common origin beyond the water snowline in the protosolar accretion disk, and thus contain a significant component of matter with a volatility similar to or less than water, preserved since the epoch of formation of the planets. Detailed compositions of icy body surfaces are routinely derived using infrared spectroscopy (obtained by ground-based observations as well as in-situ missions), showing the presence of volatiles typically dominated by H$_2$O and CO$_2$ \citep{cal95,cru10}. Less abundant ices such as N$_2$, O$_2$, H$_2$O$_2$, CH$_4$, NH$_3$, CH$_3$OH and sulphur-bearing compounds can also be identified \citep[\emph{e.g.},][]{car09,cru15}, often linked to a complex interplay between surface, subsurface, and atmospheric chemical processes over the lifetimes of these objects. Although the internal (bulk) compositions of icy bodies are more difficult to determine, the fundamental rock-to-ice ratio is often inferred based on their bulk densities, combined with knowledge of the minerals found in terrestrial, meteoritic and lunar samples.  

Europa is one of the four large moons of Jupiter, with a radius of 1560 km. It's density is relatively high among the icy bodies, with an interior composed primarily of rock and $\sim10$\% water, with an outer water ice shell, the thickness of which is uncertain but is typically believed to be in the range of a few to a few tens of kilometers \citep{sch09}. H$_2$O ice was first detected by \citet{kui57} using the McDonald observatory, but much of our recent knowledge of Europa and the nature of its interior stems from the Galileo space probe, which orbited Jupiter from 1995 to 2003 \citep{ale09}. 

Advances in our knowledge of minor icy body interiors have demonstrated the possible presence of subsurface oceans throughout the Solar System, from the asteroid belt (Ceres), to the icy moons of the outer planets, to Pluto \citep{nim16,she18}. Subsurface oceans have come to the fore as primary targets in the search for extraterrestrial life \citep{hen19}, but in-situ survey missions are costly and difficult. Plumes linking the subsurface with the atmosphere provide a potential means for probing the compositions of these liquid reservoirs, and open up the possibility of remote habitability studies.

Europa provides a unique laboratory for studying physical and chemical interactions between the atmosphere, surface and (sub-surface) ocean of an icy world. With a large reservoir of liquid water, a supply of energy from surface chemistry and interior (tidal) stress, and the availability of carbon for organic chemistry \citep{pie02}, Europa is widely recognized as a prime target in the search for life outside Earth.

Measurements of the surface ice and ocean chemical compositions are key to determining habitability \citep{chy00}, but our knowledge of Europa's full chemical inventory is far from complete.  While the surface and atmospheric compositions are known from remote sensing and spacecraft flybys, our knowledge of the subsurface ocean composition is comparatively lacking. The recent discovery of possible atmospheric plumes emanating from Europa's subsurface provides an exciting, and potentially revolutionary opportunity for investigating the sub-surface ocean properties. 

The first evidence for plumes on Europa was obtained by \citet{rot14}, who measured excess O and H emission at altitudes up to 200~km above the south-western limb, using the Hubble Space Telescope (HST). This was followed by the detection of UV limb opacity enhancements \citep{spa16}, consistent with spatially isolated surface release of gas/aerosols. Both techniques revealed (with $\approx$4-5$\sigma$ confidence), the presence of intermittent plume activity from around the same latitude in the vicinity of Europa's south pole, yet at differing longitudes, with a total estimated number of water molecules in the plume of $n_{\rm H_2O}= $1--2$\times10^{32}$. Further evidence for possible plume activity was found upon reanalysis of Galileo magnetometer and plasma wave data \citep{jia18}. The first spectroscopic detection of H$_2$O outgassing from Europa was by \citet{pag20} using Keck NIRSPEC, who reported evidence for mostly quiescent activity on 17 nights, except for one night where they measured at $n_{\rm H_2O}\sim7\times10^{31}$. Recent James Webb Space Telescope (JWST) observations by \cite{vil23} found no evidence for active plumes, indicating that any present-day activity must be localized and weak; robust confirmation of the initial HST plume results also remains challenging \citep{rot20}. Further characterization of the Europa plumes through ground-based spectroscopic observations has the potential to provide crucial new insights and additional context in preparation for the in-situ investigations to be performed by NASA's upcoming Europa Clipper mission.

\section{Observations}

We carried out observations of Europa using the Atacama Large Millimeter/submillimeter Array (ALMA) on four dates during Cycle 7, using between 43--46 antennas of the main (12 m) array, with the Band 7 receiver. The targeted spectral lines included HCN ($J=4-3$; 354.505~GHz), H$_2$CO ($J_{K_a,K_c}=5_{1,5}-4_{1,4}$; 351.769~GHz), SO$_2$ ($J_{K_a,K_c}=5_{3,3}-4_{2,2}$; 351.257~GHz), observed at 122~kHz resolution, and CH$_3$OH ($J_K = 4_0 - 3_{-1}\ E$; 350.688~GHz), observed at 61 kHz resolution (see Table \ref{tab:obs}).  Each spectral line observation was for approximately 40 minutes on Europa (on each date), and also included a dedicated continuum window of 2 GHz bandwidth, centered at around 353~GHz. All observations captured Europa's day side (with a solar phase angle of 11$^{\circ}$). The second observation was of Europa's trailing hemisphere (with respect to its orbital velocity around Jupiter), while the other three observations were of the leading hemisphere. Since Europa is tidally locked to Jupiter, the set of four observations (obtained at four different orbital phases), cover four different sub-observer longitudes (Table \ref{tab:obs}).

Spectral line data were reduced in CASA \citep{cas22} using standard pipeline scripts supplied by the Joint ALMA Observatory, including data flagging, flux, bandpass and phase calibration. Additional self-calibration procedures were applied to the continuum data, in order to improve the image fidelity and signal-to-noise ratio, and were performed using the line-free channels in all spectral windows. A model of Europa's estimated brightness temperature distribution in Band 7 (based on prior ALMA observations -- see \citealp{thelen_24a}) was generated as the starting point for self-calibration, and the subsequent iterative imaging and calibration process followed standard procedures as applied to previous observations of the Galilean satellites (see, for example, \citealp{de_kleer_21a, camarca_23, thelen_24a}, and references therein). Cleaning (deconvolution of the point spread function) was performed using the H{\"o}gbom algorithm with Briggs weighting (robust = 0.5), a pixel size of $0.02''$, a threshold of twice the continuum-subtracted spectral RMS and a $1.1''$-diameter circular mask centered on Europa. At the time of observation, Europa was at a distance of 4.6--5.2 au, which corresponds to an angular diameter of 0.8--0.9''. The angular resolution (FWHM of the Gaussian restoring beam) of the resulting image cubes, and RMS noise levels of the H$_2$CO spectral windows, are given in Table \ref{tab:obs}.

\begin{table*}[h!]
\centering
\caption{ALMA Observations of Europa \label{tab:obs}}
\begin{tabular}{crclcc}
\hline\hline
Observation Time & Long. ($^{\circ}W$) & $\Delta$ (au) & Target Species & Ang. Res. ($''$) & RMS (K)\\
\hline
2021-05-03 UT10:27 & 52 & 5.2 & HCN/H$_2$CO/SO$_2$ & $0.169 \times 0.147$ & 1.8\\ 
2021-05-09 UT10:13 & 298& 5.1 & HCN/H$_2$CO/SO$_2$ & $0.184 \times 0.152$ & 2.6\\
2021-06-08 UT11:19 & 102& 4.7 & H$_2$CO/CH$_3$OH & $0.167 \times 0.114$ & 5.0\\
2021-06-11 UT09:20 & 38& 4.6 & H$_2$CO/CH$_3$OH & $0.241 \times 0.146$ & 3.8\\
\hline
\multicolumn{6}{l}{Note --- RMS is the ($1\sigma$) spectral noise of the H$_2$CO window.}

\end{tabular}    
\end{table*}

\section{Results}

\begin{figure*}
\vspace{-5mm}
    \centering
    \includegraphics[width=0.8\textwidth]{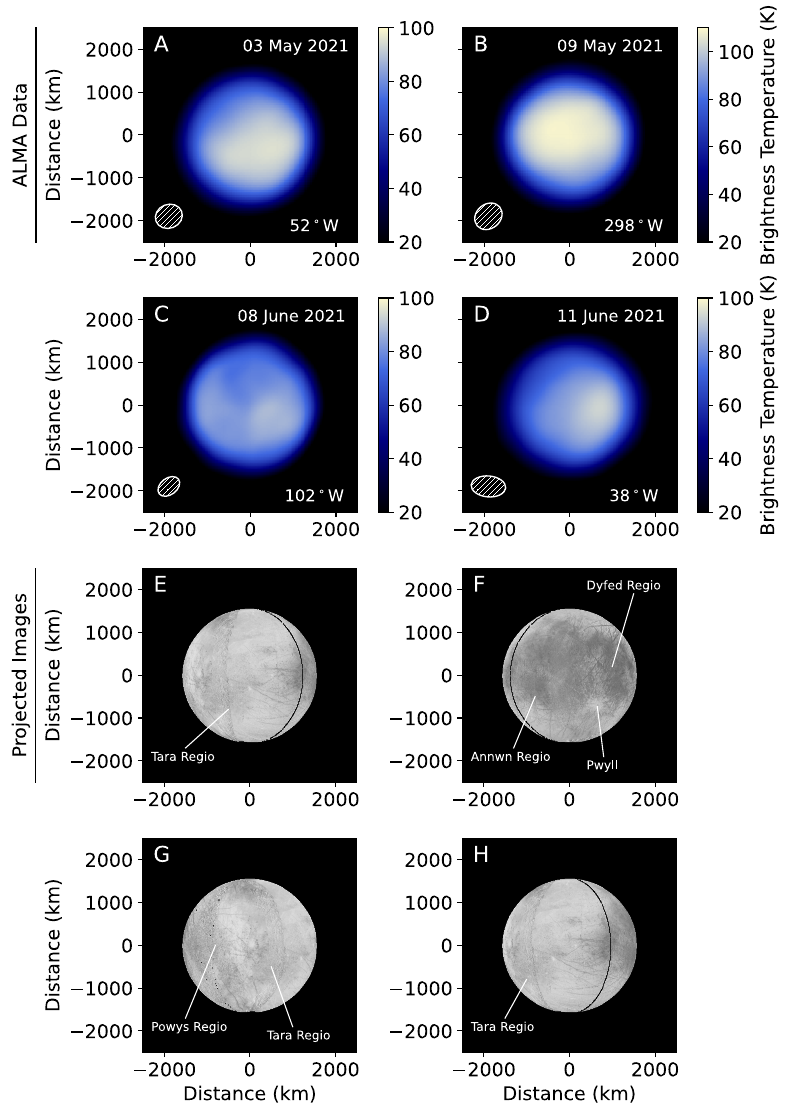}
    \caption{(A--D) ALMA Band 7 (0.9 mm) continuum brightness temperature maps of Europa on the four observing dates (rotated in the plane of the sky to align the pole with celestial north). Distances are projected in the plane of the sky. The sub-observer longitude and ALMA beam size are indicated in the lower part of each panel. (E--H) Composite projected optical images of Europa's surface from Galileo and Voyager optical data appropriate for the observing geometry of panels A--D (obtained from astrogeology.usgs.gov/search). The approximate locations of various surface features, including `regiones' and the Pwyll crater, are marked. Black line denotes 0$^\circ$W longitude.}
    \label{fig:maps}
\end{figure*}

For the four observation dates, ALMA continuum maps of Europa at 0.9 mm are shown in Figure \ref{fig:maps}, along with visual representations of Europa's surface derived from Galileo and Voyager optical imaging. The sub-mm and optical images have been rotated in the plane of the sky to align Europa's polar axis with celestial north.  Spatial structure in Europa's sub-mm brightness temperature distribution is evident, and may be interpreted as due to the combined effects of solar heating, (sub-)surface thermal emission, and surface properties (albedo, porosity, roughness, composition) of the exterior ice shell. In general, there is an apparent anti-correlation between the sub-mm (thermal) intensity and the optical brightness (albedo), which is especially evident by comparison of panel B with panel F and panel C with G. Europa's large abundance of surface H$_2$O ice results in a high average albedo (0.62; \citealt{bur83}), but the reflectivity is reduced in the presence of salts, minerals and radiolysis products \citep{dal12,bro13}. As noted by \citet{thelen_24a}, areas of low sub-mm brightness temperature may be associated with areas of either relatively high water ice content and/or lower abundances of hydrated minerals. Panel B provides a near face-on view of Europa's trailing hemisphere (the other three panels are centered primarily on the leading hemisphere). Prior ALMA observations have shown the trailing hemisphere to be consistently warmer than the leading hemisphere \citep[\emph{e.g.},][]{thelen_24a}. This is believed to be related to bombardment of the trailing hemisphere by ions accelerated along the Europa orbit by Jupiter's intense magnetosphere, which leads to a reduced abundance of pure water ice on the surface (due to sputtering and radiolysis), and results in an associated darkening at optical wavelengths.  

\begin{figure*}[h!]
    \centering
    \includegraphics[width=0.7\textwidth]{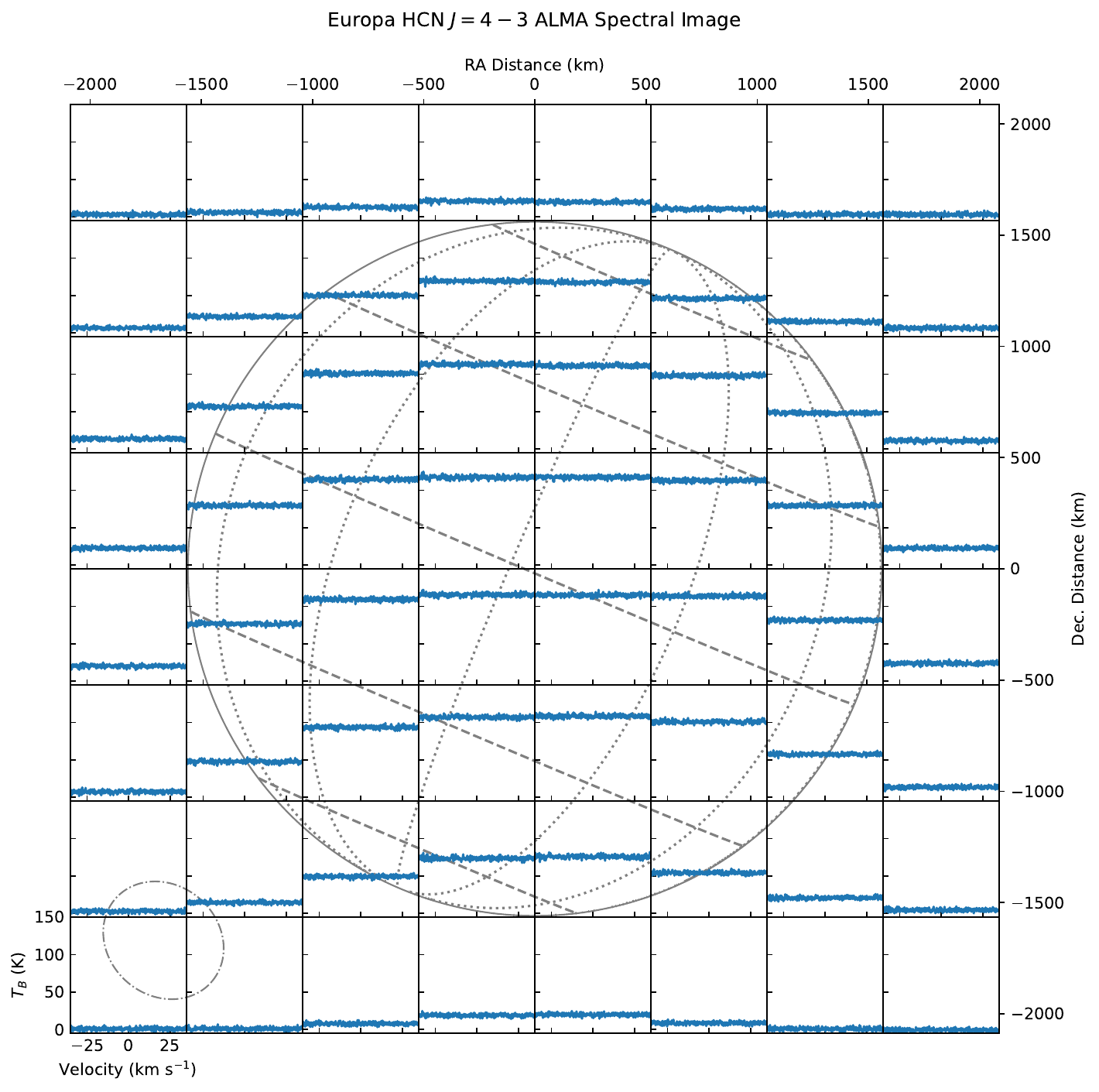}
    \caption{ALMA HCN ($J=4-3$) spectra of Europa (blue) averaged within $0.14''\times0.14''$ square regions, overlaid on a wireframe grid showing Europa's size and orientation at the time of observation (2021-05-09). The ALMA beam FWHM is indicated lower left (dot-dashed ellipse). Upper right axes show physical distances in the plane of the sky, while lower left axes show the spectral units for each sub-panel. The ALMA data for all species, on all epochs, are available for download from the Zenodo repository \citep{cor24}.}
    \label{fig:alma}
\end{figure*}

Panel G shows the presence of a bright, ice-rich band dividing two chaos regiones (Tara Regio on the right, and Powys Regio on the left), which appears relatively dark in the sub-mm (panel C). The presence of impact craters and their associated surrounding (optically bright) ejecta blankets also result in apparent reductions in the sub-mm brightness temperature; a faint dark spot towards the lower right of panel B appears to be spatially associated with the large (26 km-diameter) crater in Europa's southern hemisphere known as Pwyll. Detailed interpretation of the sub-mm maps is, of course, limited by the relatively coarse spatial resolution ($\sim500$ km in the plane of the sky), compared with the optical data.


A grid of ALMA HCN ($J=4-3$) spectra covering Europa's disk (at sub-observer longitude $298^{\circ}$) is shown in Figure \ref{fig:alma}. These data are from 2021-05-09, and are shown as representative of the typical data quality for all spectral lines, over all epochs; the HCN data from 2021-05-03 are very similar. The variation in continuum brightness across Europa is evident, but there is no sign of any apparent spectral lines in absorption or emission. Similarly for H$_2$CO, SO$_2$ and CH$_3$OH, there is no evidence for any spectral lines in the vicinity of Europa from these data. FITS data cubes and spectral image grid plots for all species, on all epochs, are available for download from the Zenodo repository \citep{cor24}.

\begin{figure}[t!]
    \centering
    \includegraphics[width=\columnwidth]{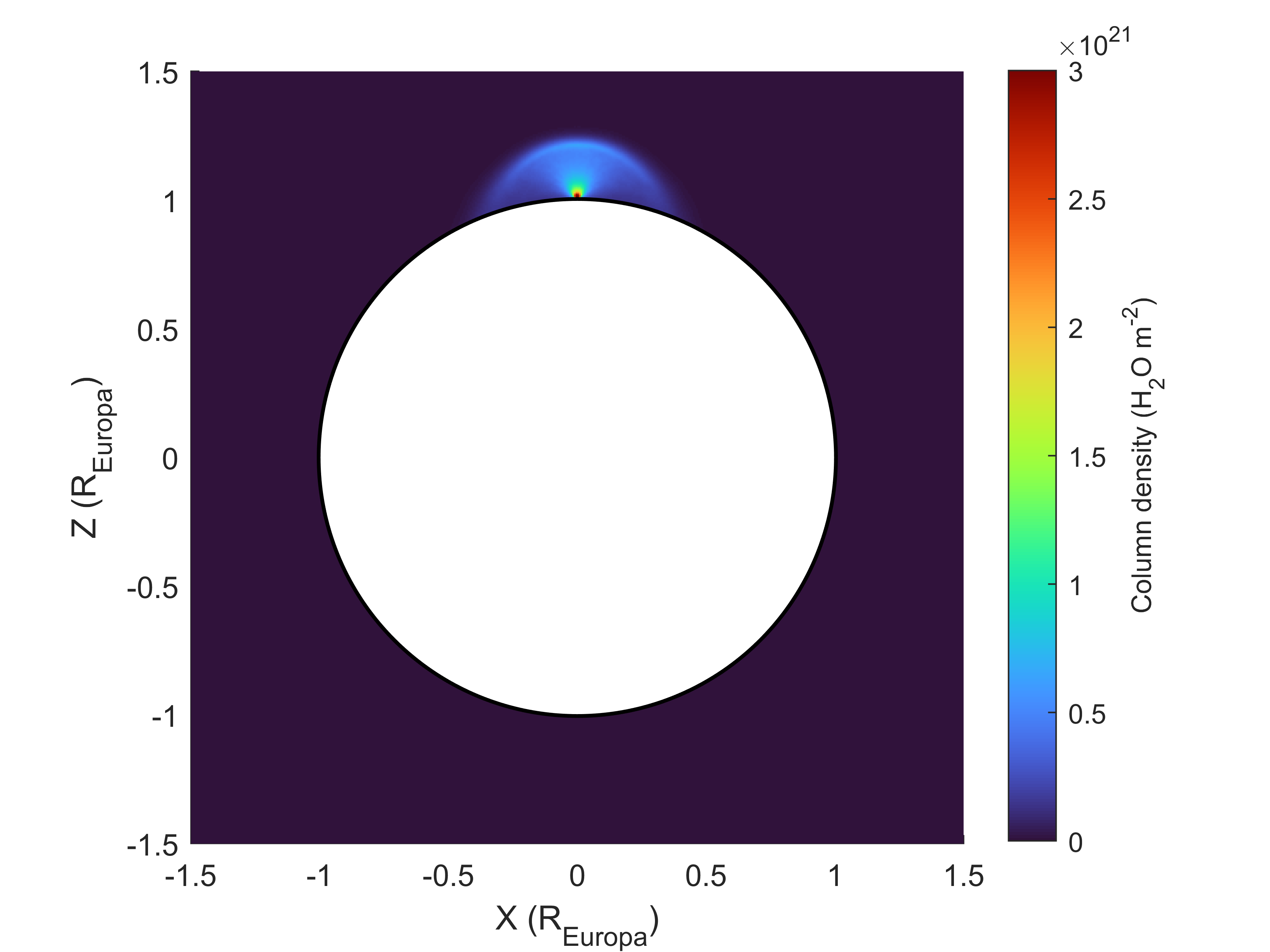}
    \caption{Column density of H$_2$O from the DSMC plume model of \citet{tse22}, emanating for the north pole of Europa, for a production rate $Q=10^{29}$~s$^{-1}$. The disk of Europa is masked with a white circle.}
    \label{fig:dsmc}
\end{figure}

To derive upper limits on the plume molecular abundances and production rates, we performed radiative transfer calculations using the LIne Modeling Engine (LIME), a 3D non-LTE excitation and Monte Carlo photon transport code \citep{bri10}. A non-LTE treatment is required due to the low density of Europa's exosphere environment, such that the molecular rotational temperature (governed by a balance of collisional and radiative processes), can deviate significantly from the gas kinetic temperature. The density, kinetic temperature and velocity distribution of the Europa plume were obtained from the 3D Direct Simulation Monte Carlo (DSMC) plume model of \citet{tse22}. DSMC is a particle-based numerical method that simulates the gas dynamics by considering collisional interactions between a large number of particles. In this case, $\sim2.5\times10^6$ particles were used. The initial plume outflow velocity was assumed to be 0.5 km\,s$^{-1}$, with a gas production rate of $Q=10^{29}$ molecules per second, a vent radius of 5~km, and initial kinetic temperature $T_{kin}=180$ K. Evolution of the resulting physical parameters was monitored across $4\times10^6$ tetrahedral cells to produce a steady state description of the plume. The total number of gas phase H$_2$O molecules in the DSMC model plume is $1.5\times10^{32}$, which is consistent with the HST observations \citep{rot14,spa16}.

The DSMC plume results were regridded onto an unstructured (Delaunay) grid for subsequent excitation calculation and raytracing in LIME. The LIME grid vertex positions were generated pseudo-randomly, with the density of points set a factor of 100 times higher inside a $30^{\circ}$ cone with its axis centered on the plume outflow axis. 50,000 grid points was found to be sufficient to adequately capture the most important features of the plume structure, including the source and canopy.  The solid surface of Europa was modeled using a dayside surface temperature of 120 K \citep{teo17} and an arbitrarily high gas density of $10^{30}$~m$^{-3}$, with gas-to-dust ratio of 100, to simulate an optically thick, black-body continuum emanating from the surface. The presence of this 120~K thermal radiation field introduces additional stimulated absorption, which is accounted for by LIME in the molecular rotational excitation calculation. Collisional excitation rates between HCN and H$_2$O were obtained from \citet{dub19}, while for the other molecules, collision rates were assumed to be the same as for H$_2$, and were obtained from the LAMDA database \citep{van20}. The model geometry was set to place the plume origin at a radial emission angle of $45^{\circ}$ with respect to the line of sight, which represents the statistical average of the possible emission angles considering a random plume location on Europa's observer-facing hemisphere.

\begin{figure*}
    \centering
    \includegraphics[width=0.7\textwidth]{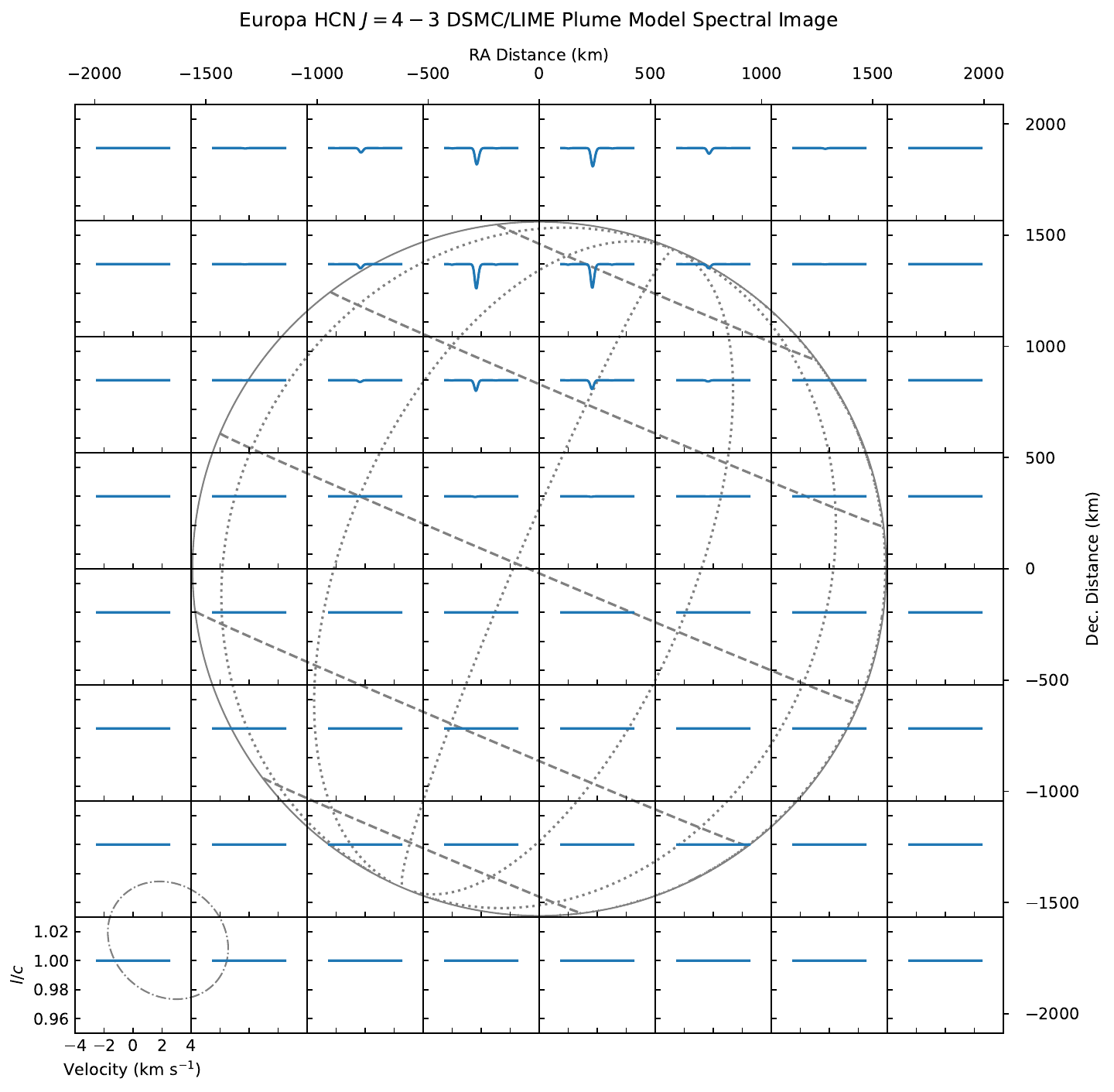}
    \caption{Radiative transfer model of a Europa HCN plume based on the DSMC model of \citet{tse22}, convolved to the spatial resolution of the ALMA observations and plotted as a line-to-continuum ratio ($l/c$). The plume emission angle was set to $45^{\circ}$ north of the sub-observer point, with an HCN abundance of $10^{-5}$. Spectra (shown in blue) have been averaged within $0.14''\times0.14''$ square regions for comparison with Figure \ref{fig:alma}.}
    \label{fig:model}
\end{figure*}

Raytracing of the plume model was performed on a Cartesian grid with pixel sizes $0.01''$ and a spectral resolution of 0.05~km\,s$^{-1}$. Finally, the model results were convolved with the ALMA beam size for comparison with observations (see Fig. \ref{fig:model}). The spectral model line intensity ($l$) was divided by the continuum brightness ($c$) to produce the model line-to-continuum ratio ($l/c$)$_{model}$. This was compared with the observed integrated RMS noise level ($W=\int(\sigma(l)/c)_{obs}dv$, where the integral is over the model line full-width of 0.7~km\,s$^{-1}$), to determine $3\sigma$ upper limits on the molecular abundances and production rates, which are given for each species and observation date in Table \ref{tab:results}.

\begin{table*}[ht!]
\centering
\caption{Molecular abundance and production rate $3\sigma$ upper limits \label{tab:results}}
\begin{tabular}{lcllc}
\hline\hline
Species & Observation Date & $W$ (m\,s$^{-1}$) & Abund. (\%) & Q (s$^{-1}$)\\
\hline
H$_2$CO & 2021-05-03&  6.2 & $<0.018$ & $<1.8\times10^{25}$ \\
        & 2021-05-09&  6.4 & $<0.016$ & $<1.6\times10^{25}$ \\
        & 2021-06-08&  14 &  $<0.030$ & $<3.0\times10^{25}$ \\
        & 2021-06-11& 10 &   $<0.021$ & $<2.1\times10^{25}$ \\
HCN  & 2021-05-03& 8.1 &     $<0.004$ & $<4.4\times10^{24}$ \\
     & 2021-05-09& 8.0 &     $<0.004$ & $<3.8\times10^{24}$ \\
SO$_2$ & 2021-05-03& 8.8 &   $<0.15 $ & $<1.5\times10^{26}$ \\
       & 2021-05-09& 6.4 &   $<0.09$ & $<9.0\times10^{25}$ \\
CH$_3$OH & 2021-06-08& 12 &  $<1.16$  & $<1.2\times10^{27}$ \\
       & 2021-06-11&  9.1 &  $<0.86$  & $<8.6\times10^{26}$ \\
\hline

\end{tabular}    
\end{table*}

\section{Discussion}

Despite near-complete coverage of both Europa's leading and trailing hemispheres, we find no evidence for gas phase molecular absorption or emission in our ALMA data. Assuming a production rate for H$_2$O molecules in the plume of $10^{29}$~s$^{-1}$, our strictest ($3\sigma$) upper limit on the H$_2$CO abundance is $<0.016$\%, which is more than an order of magnitude less than observed in the Enceladus plume by the Cassini spacecraft’s Ion and Neutral Mass Spectrometer \citep[INMS;][]{wai09}.  Our HCN upper limit of $<0.004$\% is more than two orders of magnitude more restrictive than the Enceladus INMS measurement of $<0.7$\%. For the CH$_3$OH abundance, on the other hand, our ALMA upper limit of $<0.86$\% would not have been sensitive enough to detect this molecule at the Enceladus plume abundance of 0.02\%. 

In the presence of a plume, the lower H$_2$CO abundance observed for Europa implies a different plume composition compared with Enceladus. The similarity of the Enceladus plume H$_2$CO abundance to that found in comets was noted by \citet{wai09}, and, together with the abundances for other cometary molecules in the plume, implies the presence of subsurface ice sublimation products outgassing from Enceladus. The lack of H$_2$CO in the Europa plume could imply that the latter is dominated instead by outgassing from an aqueous (ocean) environment. 

Interpretation of these abundance upper limits should, however, account for the intermittent nature of the Europa plume, as identified by \citet{spa17} and \citet{pag20}. If our ALMA observations were all conducted during a low (or zero) activity plume phase, then our abundance upper limits would be correspondingly larger. In that case, it is more useful to consider the molecular production rates ($Q$; see Table \ref{tab:results}).  The fact that the Europa plume physical structure is inferred based on DSMC simulations, whereas the actual size, velocity and kinetic temperature of the plume have not yet been accurately measured, introduces some additional uncertainties into our abundance and production rate results. In particular, a higher plume outflow velocity would result in a larger canopy and correspondingly lower molecular abundance estimates. If the plume extended over a larger area than assumed by our model, then our measurements (for a single ALMA beam) could be further underestimated.

The continuum images of Europa (Fig. \ref{fig:maps}A--D) show similar brightness temperature distributions to previous observations of Europa with ALMA \citep{trumbo_17, trumbo_18, thelen_24a}, including elevated temperatures colocated with Europa's expansive, optically-dark regiones (\emph{e.g.} Fig. \ref{fig:maps}C), and across the trailing hemisphere (Figure \ref{fig:maps}B). The localized areas of reduced brightness temperatures at northern latitudes on the leading hemisphere, and within the expansive ejecta blanket and rays surrounding Pwyll crater (Figure \ref{fig:maps}B) were identified by \citet{thelen_24a} to possess higher thermal inertia than the surrounding terrain. A detailed physical model of Europa's (sub-)surface thermal emission (\emph{e.g.} \citealt{de_kleer_21a}) would be required for a full interpretation of the ALMA sub-mm continuum maps, to enable a search for thermal anomalies that may be associated with plume activity, but such an analysis is beyond the scope of the present work. 

Recent infrared spectral mapping observations by the JWST identified enhanced concentrations of CO$_2$ within Tara Regio \citep{vil23,tru23}, while an excess of sub-mm (thermal) emission was detected in this area by \citet{thelen_24a} using ALMA. This region was covered by spectral line and continuum observations on three of our ALMA observing epochs. The presence of spatially isolated thermal anomalies in Tara Regio could be explained as a result of high salinity, high porosity or low thermal inertia of the surface, or more intriguingly, could point towards unusual geological activity or hot-spots associated with plume activity or thinning of the ice shell. 

The fact that no evidence for plume outgassing was identified by our observations indicates that either the Europa plumes are intermittent, or if they were active at the time of our observations, then they contain very small (or even zero) abundances of the molecules we searched for. Additional ALMA observations, including followup searches for other molecules found in the Enceladus plumes (such as H$_2$O, NH$_3$, NaCl and KCl), will be required to better understand the composition and temporal cadence of Europa's elusive plumes. Further investigation of possible chemical differences between the Europa and Enceladus plumes would provide fundamental insights into the comparative compositions of the respective plume source materials, in order to help constrain the properties of the subsurface solid, gas or liquid reservoirs.

\section{Conclusion}

Using ALMA's unique combination of high spectral/spatial resolution and sensitivity, our observations have enabled the first dedicated search for HCN, H$_2$CO, SO$_2$ and CH$_3$OH in Europa's exosphere and plumes. No evidence was found for the presence of these molecules and strict ($3\sigma$) upper limits were derived on their abundances and production rates, by comparison with 3D plume DSMC and radiative transfer modeling. Our most restrictive plume production rate upper limits are for HCN ($<3.8\times10^{24}$~s$^{-1}$) and H$_2$CO ($<1.6\times10^{25}$~s$^{-1}$), showing that if plumes were active at the time of observation, they contained very low abundances of these molecules (as well as of SO$_2$ and CH$_3$OH). Indeed, our H$_2$CO abundance upper limit is significantly lower than measured by Cassini in the Enceladus plume, implying a possible chemical difference. Our results show that ALMA is a powerful tool in the search for outgassing from icy bodies within the Solar System, and that followup searches for other molecules at additional epochs (on Europa and other icy bodies) are justified. 

\section{Acknowledgements}

This work was supported by the National Science Foundation under grant No. AST-2009253 and by NASA’s Planetary Science Division Internal Scientist Funding Program through the Fundamental Laboratory Research work package (FLaRe). This work makes use of ALMA data set ADS/JAO.ALMA\#2018.1.01114.S. ALMA is a partnership of ESO, NSF, NINS, NRC, NSC and ASIAA, in cooperation with the Republic of Chile. The Joint ALMA Observatory is operated by ESO, AUI/NRAO and NAOJ.

\bibliographystyle{iaulike}
\bibliography{refs}{}

\end{document}